\documentclass[11pt,manuscript]{aastex}
\begin{document}
%\graphicspath{{Figures/}}

%\newcommand{\arcmin}{$^{\prime}$}
%\newcommand{\arcsec}{$^{\prime\prime}$}
%\newcommand{\arcmin}{$\rm arcmin$}
%\newcommand{\arcsec}{$\rm arcsec$}
\newcommand{\degree}{$^{\circ}$}
\newcommand{\ve}{$v_{\rm e}$}

%% If you wish, you may supply running head information, although
%% this information may be modified by the editorial offices.
%% The left head contains a list of authors,
%% usually a maximum of three (otherwise use et al.).  The right
%% head is a modified title of up to roughly 44 characters.
%% Running heads will not print in the manuscript style.

%\shorttitle{Large Fragments in P/2010 A2}
%\shortauthors{Agarwal, Jewitt, ...}

%\begin{document}

\title{Dynamics of Large Fragments in the Tail of Active Asteroid P/2010 A2}

%\author{Jessica Agarwal$^1$ and David Jewitt$^2$}
%\affil{$^1$ Max Planck Institute for Solar System Research, Max-Planck-Str. 2,
%37191 Katlenburg-Lindau, Germany\\
%$^2$Department of Earth and Space Sciences and Department of Physics and Astronomy, University of California at Los Angeles, 595 Charles Young Drive East, Los Angeles, CA 90095-1567}
%\email{agarwal@mps.mpg.de}

\author{Jessica Agarwal\altaffilmark{1}}
\affil{Max Planck Institute for Solar System Research, 37191 Katlenburg-Lindau, Germany}
\email{agarwal@mps.mpg.de}

\author{David Jewitt\altaffilmark{2}}
\affil{Department of Earth and Space Sciences and Department of Physics and Astronomy, University of California at Los Angeles, Los Angeles, CA 90095-1567, USA}

\author{Harold Weaver\altaffilmark{3}}
\affil{The Johns Hopkins University Applied Physics Laboratory, Johns Hopkins University, Laurel, Maryland 20723, USA}

%\date{\textbf{DRAFT 2013 Feb 18}}

\begin{abstract}
We examine the motions of large fragments at the head of the dust tail of active asteroid
P/2010 A2.  In previous work we showed that these fragments were ejected from the primary nucleus in early 2009, either following a hypervelocity impact or by rotationally induced break-up.  Here, we follow their positions through a series of Hubble Space Telescope images taken during
the first half of 2010. The orbital evolution of each fragment allows us to constrain its velocity relative
to the main nucleus after leaving its sphere of gravitational influence. 
We find that the fragments constituting a prominent $X$-shaped tail feature were emitted in a direction opposite to the motion of the asteroid and towards the south of its orbital plane. Derived emission velocities of these primary fragments range between 0.02 and 0.3 m s$^{-1}$, comparable to the $\sim$0.08 m s$^{-1}$ gravitational escape speed from the nucleus.  Their sizes are on the order of decimeters or larger. We obtain the best fits to our data with ejection velocity vectors lying in a plane that includes the nucleus. This may suggest that the cause of the disruption of P/2010 A2 is rotational break-up.
\end{abstract}


\section{Introduction}
The active asteroid P/2010 A2 was discovered in January 2010 by the LINEAR Sky Survey \citep{kadota-blythe2010}. This inner-belt asteroid displayed a long, narrow tail resulting from the ejection of dust. High resolution
imaging with the Hubble Space Telescope (HST) showed that the dust in the tail seemed to emerge
from a bright $X$-shaped pattern of large dust grains rather than from the main nucleus, which
appeared almost detached from the tail \citep[Jewitt et al. 2010, hereafter referred to as][]{jewitt-weaver2010}.
%, hereafter Paper 1). 
Dynamical analysis of the tail's position angle on the sky
revealed that the dust was emitted from the nucleus during a very short time span about nine months
before the discovery of P/2010 A2. We therefore concluded that, in 2009 February or March,
P/2010 A2 either was impacted by a second asteroid or perhaps disrupted due to rotational break-up.  The disrupted P/2010 A2 went unobserved from this initial event to (pre-discovery) observations taken in 2009 November 22 \citep{jewitt-stuart2011}.
The basic features and possible interpretation of P/2010 A2 have been confirmed by a number of independent investigators \citep {jewitt-weaver2010, snodgrass-tubiana2010, hainaut-kleyna2012, kim-ishiguro2012, kleyna-hainaut2013}.  In particular, \citet{kleyna-hainaut2013} have presented a detailed impact model that purports to fit many of the observed properties of the tail structure.  

Whether by impact, rotational disruption, or another process, observations of P/2010 A2 promise to throw light on the physics of asteroid disintegration, a process that has previously gone unobserved.  Ultimately, this will be important both to an understanding of the size distribution of the sub-kilometer asteroids and to the process of dust and debris production in the solar system, including the formation of meteoroid streams (like the Geminids) which have an asteroidal parent.  

In the following, we examine the motions of large particles that form the $X$ at the head of the dust tail. We constrain the ejection velocities and sizes of these bodies by comparing a series of HST images to the simulated trajectories of test particles ejected from the nucleus with variable initial parameters. The data and model are described in Section~\ref{sec:data_model}, and the results are presented in Section~\ref{sec:results}. The discussion in Section~\ref{sec:discussion} addresses the implications of our findings for the reconstruction of the ejection mechanism. We also compare our results to additional observations of P/2010 A2 that were not part of the data set on which our analysis is based. Finally, we outline how the $X$-pattern could have formed as a consequence of fast rotation, and describe how future observations might help to discriminate between collisions and rotational break-up of asteroids. Our results are summarised in Section~\ref{sec:summary}.

\section{Data and Model}
\label{sec:data_model}
A majority of the tail features are invisible in lower resolution data obtained from the ground.  Accordingly, we use exclusively high resolution imaging data obtained from the HST. These observations were first published in \citet{jewitt-weaver2010}, and their dates and geometrical circumstances are listed in Table~\ref{tab:obs_geometry}.

We study here the $X$-shaped structure at the head of the tail, close to the nucleus. The early images show that it is formed by two bright arcs that intersect. From 2010 January to May, this structure became more compressed in the north-south direction, and the tail as a whole became narrower, even in images corrected for the steadily increasing geocentric distance (c.f.~Table~\ref{tab:obs_geometry}). This progressive contraction occurs because the images were made more than a quarter of the orbital period after the emission, and the dust was again approaching the orbital plane of the nucleus. However, the overall structure of an $X$ with several particularly bright spots remained stable. In particular, and as already noted in \citet{jewitt-weaver2010}, the morphology of the $X$ did not change as the Earth moved from one side of the orbital plane to the other, showing that this feature is extended perpendicular to the plane.  

Our measurement strategy is to identify features in the tail and to follow their positions through the HST image series.   The features were identified visually, a task made difficult by their intrinsic faintness and the spatial complexity of the structured tail.  We assume that each feature refers to the same material in each image. For features at and near the ends of the $X$ structure, this seems a good assumption, while for the point of intersection of the two arms we cannot be certain that it is more than a projection into the plane of the sky. Nevertheless, we measured also the path of this point and studied it in the same way as the points at the ends.

The regions we study are shown in Fig.~\ref{fig:region_nomenclature}.  We label features on the arm extending from the nucleus to the lower right in Fig.~\ref{fig:region_nomenclature} by ``A'' while features on the other arm are labeled ``B''. The position of each feature is measured where possible (Fig.~\ref{fig:region_paths}) and used to compute an ephemeris.  The linear resolution of the data degrades with increasing geocentric distance, so that the features generally become less distinct from 2010 January to May, even though the intrinsic angular resolution of the HST data is stable. The last observation of the HST images series (dating from UT 2010 May 29) is not considered in this analysis, because it has too low signal-to-noise ratio and resolution.

We assume that the material in the circled regions was separated from the main nucleus at a single moment in time as found by the synchrone analysis in
 \citet{jewitt-weaver2010}. We study three dates of emission covering the interval given in \citet{jewitt-weaver2010}, namely UT 2009 February 9, March 2, and March 23. After leaving the nucleus' sphere of gravitational influence (Hill sphere) the motions of the ejecta are determined by their velocity vector on leaving the Hill sphere, and by their radiation pressure coefficient $\beta$, which depends on the physical properties and size of the fragments and is equal to the ratio of the accelerations due to radiation pressure and solar gravity \citep{burns-lamy1979}. The radius of the Hill sphere of P/2010 A2, for an assumed density of 3000 kg m$^{-3}$ and radius of 60\,m, is 23\,km, which is below the pixel scale of our observations (cf.~Table\ref{tab:obs_geometry}).

The three velocity components, $v_x$, $v_y$, $v_z$, and $\beta$ of the material are the free parameters to be derived from the observed path of each region. 
The velocity components are defined as follows: the $v_y$-direction is parallel to the orbital velocity vector of the nucleus at the time of emission, $v_z$ is perpendicular to the orbital plane of the nucleus, and $v_x$ is perpendicular to both, pointing away from the Sun.

During the observations, all considered regions must have been located south of the orbital plane, because they appeared south of the projected orbit both before and after Earth crossed the orbital plane of P/2010 A2 in February 2010. Since the observations took place less than half an orbital period after the emission, we can infer that the fragments left the Hill sphere of the asteroid towards the south of the orbit. 

For each emission date, we calculate the trajectories for $\approx 10^8$ test particles with $0 \leq v_l \leq$ 1 m s$^{-1}$ in steps of 0.01 m s$^{-1}$ ($l=x,y,z$), and $0 \leq \beta \leq 10^{-4} $ in steps of $10^{-6}$. Having determined the approximate region in parameter space that contains possible solutions, we refine our grid by a factor of ten.

For each parameter set, $j$, we calculate the position of the test particle at the observation dates $i$, $(x_i^j, y_i^j)$, and the distance $d_i^j$ between the center of the observed region and the test particle: $(d_i^{j,k})^2 = (x_i^j-\xi_i^{k})^2 + (y_i^j-\eta_i^{k})^2$, where ($\xi_i^{k},\eta_i^{k}$) describes the position of the region $k$ in the HST image $i$. Ultimately, we obtain for each test particle, $j$, and for each region, $k$, the quantity $D_{j,k}=(\sum_{i=1}^7 (d_i^{j,k})^2)^{1/2}$, which we seek to minimise. 
% $D_{j,k}$ is calculated in lstsq.f
We consider all parameter sets $j$ as acceptable solutions for the region $k$ that fulfill the condition $d_i^{j,k} < \epsilon_i^k$ for all observation dates $i$, where $\epsilon_i^k$ is the radius of the circular region $k$ in image $i$ (c.f. Fig.~\ref{fig:region_paths}). The best solution is characterised by having the minimum $D_{j,k}$.
% In lstsq_maxdist.f, only parameter sets with $d_i^{j,k} < \epsilon_i^k$ for all $i$ are printed.

\section{Results}
\label{sec:results}
For each region identified in Fig.~\ref{fig:region_nomenclature}, the possible solutions represent a continuous region in the $v_x$-$v_y$-$v_z$-$\beta$-space.
Fig.~\ref{fig:vspace_regions_all} shows
for a given date of emission, the regions in velocity space that correspond to possible solutions for each of the image regions identified in Fig.~\ref{fig:region_nomenclature}. The figures show the projections of these allowed regions in three-dimensional velocity space to the $v_x$, $v_y$- and $v_x$, $v_z$-planes. The best fit (minimum $D_{j,k}$) is represented by a cross. 

The allowable regions in velocity space have the shapes of half Zeppelin-like, prolate spheroids with axis ratios of about 16:1. Surfaces of constant fit-quality ($D_{j,k}$) correspond to half-spheroids nested into each other. Allowed parameter sets for constant values of $\beta$ lie close to parallel planes in velocity space, perpendicular to the $v_x$-$v_y$-plane (Fig.~\ref{fig:beta_sheets_example}). There is only a negligible correlation between $\beta$ and $v_z$, but all allowed solutions for a given $\beta$ are characterised by a roughly linear relationship between the $v_x$ and $v_y$ components: 

\begin{equation}
v_y= m v_x + k \, \beta.
\label{eq:relation_v_beta}
\end{equation} 

The approximately circular cross-section of the allowed regions in velocity space results from the circular shape of the regions we study in image coordinates. The extent along the long axis of the spheroid is due to the fact that different combinations of ejection velocity and radiation pressure parameter give the same total energy of the particle, which is equivalent to the same orbital period and therefore to the same projected distance from the nucleus \citep[e.g.,][]{mueller-green2001}.

The best fits are achieved with values of $\beta<2 \times 10^{-7}$, while less-likely but still formally acceptable solutions have $\beta<2 \times 10^{-5}$. The best fitting solutions have $D_{j,k}$ on the order of 0.1 to 0.4 arcsec (depending on region and ejection time), and the barely acceptable solutions with lowest fit quality have $D_{j,k}$ between 0.3 and 1 arcsec.

The dust in the regions we study has been emitted in directions opposite to the motion of the comet (negative $v_y$) and to the south of the orbital plane (negative $v_z$). The best fitting solutions have a $v_x$-component pointing away from the Sun, but ejection towards the inside of the orbit is possible for some fragments at higher values of $\beta$ and lower fit quality.

Qualitatively, the results are similar for all studied emission dates, but the fit quality is highest for early emission (9 February 2009). The regions with strong out-of-plane velocity components could not be fitted with later emission dates at all, consistent with the dating of the disruption in \citet{jewitt-weaver2010} and \citet{snodgrass-tubiana2010}. 

The vertical emission velocity components ($v_z)$ range from 0 to 0.15 m s$^{-1}$, and are very specific for each region. The distance of a region from the projected orbit in the HST images is almost exclusively determined by the $v_z$-component of the contained dust. 

The ejection velocity components in the orbital plane ($v_x$ and $v_y$) range between 0 and 0.3 m s$^{-1}$, and the direction of emission is correlated with $\beta$ (cf. Fig.~\ref{fig:beta_sheets_example}). The best fitting solutions for different image regions and a given date of emission lie close to a plane that is perpendicular to the $v_x$-$v_y$-plane, but not parallel to the planes of constant $\beta$. For emission in early February, the plane of best fits includes the nucleus. Within the plane of best fits, the ejection velocity vectors of the regions we study are distributed in a pattern similar to that of the observed regions on the sky. Thus we see the $X$ also in velocity space (see the $v_x$-$v_z$-projections in Fig.~\ref{fig:vspace_regions_all}).

\section{Discussion}
\label{sec:discussion}

\subsection{Ejection pattern}
\label{subsec:ejection}
In the following we discuss the motion of material in the $X$ at the head of the tail of P/2010 A2. Farther away from the nucleus, the tail consists of smaller particles driven away by radiation pressure \citep{jewitt-weaver2010, snodgrass-tubiana2010}. The dynamics of these particles is addressed in Section~\ref{subsec:model_images}.

The escape speed from the surface of a spherical nucleus of 60 m radius and a density of 3000 kg/m$^3$ is 0.08 m s$^{-1}$. The uncertainty of the escape speed from P/2010 A2 is at least a factor of $\sqrt2$, due to the uncertainty of the albedo and therefore radius \citep{jewitt-weaver2010, jewitt-ishiguro2013}. 

By escape speed we refer to the speed {\em relative to the nucleus center} required at surface level to leave the gravitational influence of the nucleus. 
On a rotating body, the surface material has some kinetic energy due to the rotation. The initial speed {\it relative to the surface} required to escape the gravitational influence of the rotating nucleus depends on the latitude from which material is launched. It is smallest at the equator, while at the poles no rotation effect is seen and the required speed relative to the surface is the same as the speed relative to the nucleus center. (To make use of this effect, rockets on Earth are preferentially launched close to the equator.)
In the case of critical rotation, the required speed relative to the surface at the equator is zero. Viewed in an inertial frame, the rotation velocity at the surface (i.\,e. the initial velocity of the grain) is then equal to the escape speed from the non-rotating nucleus, which defines the critical rotation period. Analysing only the motion of the dust, it is not possible to distinguish between material separated with zero relative speed from a critically rotating body or ejected with escape speed tangentially to the surface from a non-rotating body. Note that this definition of the critical period does not take into account any material strength. Since we observe speeds that exceed the nucleus escape speed, the material at the surface must have been attached to the nucleus by cohesive forces prior to breaking away from it.

The ejection speeds we infer for the material in the $X$ are of the same order of magnitude as the escape speed. We find the lowest ejection speed for the regions very close to the nucleus (0.02 m s$^{-1}$ for region A1 and 0.04 m s$^{-1}$ for A2).  We note here that the ejection velocity in our model means the velocity on leaving the sphere of gravitational influence of the nucleus.
At low ejection velocities, the influence of gravitational deceleration can be considerable, and we cannot approximate the velocity at the surface by the measured one. Apart from decelerating, the material may have changed direction between leaving the surface and decoupling from the gravity field of the nucleus. 

We find it remarkable that the speed distribution of the ejecta ends 
abruptly at values on the order of 0.2 m s$^{-1}$, which is about twice the escape speed from the nucleus surface. 
The lack of faster ejecta  is qualitatively consistent with both impact and rotational break-up origins. In an impact, the fast ejecta carry only a small fraction of the ejected mass and will, in any case, have escaped the field of view of HST in the $\sim$1 yr separating ejection from observation. The bulk of the mass in an impact is ejected at the lowest velocities \citep{housen-holsapple2011}, limited eventually by the need to gravitationally escape the nucleus.
If, instead, P/2010 A2 disrupted due to fast rotation, the cut-off velocity would be related only to the radius, spin rate and tensile strength of the nucleus, and no faster material is expected.

A strong indication that the ejection of material from P/2010 A2 may have been due to rotation is our finding that the best fits to our data are obtained with velocity vectors that lie close to a plane passing through the nucleus 
(cf.~Fig.~\ref{fig:vspace_regions_all}).
This plane could correspond to the equatorial plane of the nucleus, which would mean that the rotation axis lies in the orbital plane of P/2010 A2.

While the planar solution for the ejected material in the $X$ is not unique, others that fit the data require assumptions about the ejecta that are more contrived. For example, we can fit the fragment position measurements by arbitrarily assuming that there exists a correlation between the maximum fragment size and the emission direction 
(cf.~Fig.~\ref{fig:vspace_regions_all}),
but this assumption has no obvious physical basis.

\subsection{Model images and comparison to other observations}
\label{subsec:model_images}
To cross-check and illustrate our results, we calculated model images based on the best fitting parameters. We compare model images to three observations: Fig.~\ref{fig:simimage} is a comparison to the HST image taken in 2010 January 29 and published in \citet{jewitt-weaver2010}. This image is part of the data set on which our study is based (cf. Fig.~\ref{fig:region_nomenclature}, second row). Fig.~\ref{fig:ros} addresses an observation made from the Rosetta spacecraft on 2010 March 16 \citep{snodgrass-tubiana2010}, and Fig.~\ref{fig:keck} is a comparison with an image taken on 2012 October 14 with the Keck telescope \citep{jewitt-ishiguro2013}.

To generate simulated images, we calculated the positions of 10$^7$ particles, launched from the nucleus on 2009 February 9 with relative initial velocities interpolated from the best fits obtained for the selected regions. We use not only the large particles located in the $X$, but particles covering the whole size range seen in the 2010 HST image. The initial velocity does not depend on the particle size, only on the ejection direction.

The radiation pressure coefficient was taken in the range $10^{-7}< \beta < 10^{-3}$, corresponding to sizes between 0.5\,mm and 5\,m for a density of 1000 kg m$^{-3}$. The radiation pressure coefficient is distributed as ${\rm d}n / {\rm d}\beta \propto \beta^{\gamma}$, which is related to the differential size distribution ${\rm d}n / {\rm d}s \propto s^{\alpha}$ through $\gamma = -\alpha -2$ if the bulk density and optical properties of the grains are independent of their size. To generate the model images in Figs.~\ref{fig:simimage} to \ref{fig:keck}, we used $\alpha=-3.3$, inferred from the brightness slope in the distant tail in the HST images \citep{jewitt-weaver2010}.
To generate a smooth dust distribution inside the cross, we interpolated continuous initial velocities between the localised regions shown in 
Fig.~\ref{fig:vspace_regions_all} (top panels) through fitting empirical relations for $v_y (v_x)$ and $v_z(v_x)$ (all velocities in m s$^{-1}$):

\begin{eqnarray}
\label{eq:empirical_relation_v}
\nonumber v_y = -0.612 v_x + 0.015\\
v_{zA} = -12 v_x^2\\
\nonumber v_{zB} = -12 (0.135 - v_x)^2 - 0.015
\end{eqnarray}

\noindent where subscripts $A$ and $B$ refer to the two arcs.  The relation $v_z (v_x)$ is specific to each arc, while $v_y(v_x)$ is the same for both, such that all initial velocities lie in a plane (Fig.~\ref{fig:best_fit_empirical}). 

We generated uniformly distributed random values for $v_x$ in the intervals 0.02 m s$^{-1}$ $< v_x <$ 0.11 m s$^{-1}$ for arc A, and 0.045 m s$^{-1}$ $< v_x <$ 0.13 m s$^{-1}$ for arc B, and calculated the corresponding $v_y$ and $v_z$ from Eqs.~\ref{eq:empirical_relation_v}. We added a random component to this velocity vector, distributed uniformly inside a spherical volume of radius 0.004 m s$^{-1}$ in velocity space, to account for the finite extent of the regions inside the $X$. All particles in the simulated images had initial velocities distributed between the lines shown in Fig.~\ref{fig:best_fit_empirical}. Their spreading out in the tail is due to radiation pressure only. 

The simulated image in Fig.~\ref{fig:simimage} (bottom) qualitatively reproduces both the $X$ and the streaks observed in the tail of P/2010 A2, marked by arrows in Fig.~\ref{fig:simimage} (top).
Our model assumption, that ejection velocity and size are uncorrelated, allows us to reproduce the observed streaks and interpret them as enhanced numbers of ejected particles for certain ejection directions (e.g. region B3). 

Two ejection scenarios would lead to size-independent velocities:
Either, ejected decimeter and larger-sized parent fragments subsequently decayed into smaller particles which were then accelerated by radiation pressure to form the parallel streaks in the tail. 
Alternatively, all particles may have originally been ejected at the same velocity, which we would expect from a regolith surface accelerated to critical rotation speed, but also as the result from an impact. About one year after the ejection, the largest fragments would be found close to the nucleus (inside the $X$), while smaller particles would be in the tail due to the action of radiation pressure. 

Note that to generate Figs.~\ref{fig:simimage} to \ref{fig:keck} we considered only the regions identified in Fig.~\ref{fig:region_nomenclature} and interpolated between them. For this reason, Fig.~\ref{fig:simimage} does not show the dust features at the northern edge of the tail, which we have not studied. Fig.~\ref{fig:simimage} merely illustrates that the parameters we found indeed reproduce the $X$ structure we set out to explain.

Fig.~\ref{fig:ros} (right) shows an image of our model dust as seen from the point of view of the Rosetta spacecraft on 2010 March 16, the time of the observation described in \citet{snodgrass-tubiana2010} (Fig.~\ref{fig:ros}, left). While the orientation of the tail is comparable in both, the simulated tail is more narrow than the observed one. The reason may be that through picking the seven specific regions in the $X$, we capture only a subset of the dust present in the tail. The Rosetta image suggests that there may have been a cloud of dust extending parallel to the orbital plane of the nucleus.

In Fig.~\ref{fig:keck} we compare our model to an observation taken on 2012 October 14 with the Keck telescope \citep{jewitt-ishiguro2013}.
The observation showed a dust trail extending to both sides of the nucleus. In our model, most of the material that formed the $X$ in 2010 is located to the east of the nucleus in 2012 (leading the nucleus in its orbital motion), hence the model is consistent with the existence of the eastern trail in the 2012 image. As in Figs.~\ref{fig:simimage} and \ref{fig:ros}, the model trail is more narrow than observed, and the peak surface brightness is shifted with respect to the nucleus. Both aspects support the hypothesis that there was an additional diffuse component of dust not captured by our modelling of the $X$. 

\citet{kleyna-hainaut2013} offer an interpretation of the two arcs as the edge (arc A) and rim (arc B) of a hollow ejecta cone resulting from an impact. While they start from the assumption of cone-shaped ejection, our approach makes no prior assumptions about the relations between ejection direction, velocity, and grain size. On the other hand, our approach rests on the assumption that the regions identified on the $X$ contain the same material throughout the four months' image series. Provided the correctness of this assumption, our resulting ejection velocities 
(cf. Fig.~\ref{fig:vspace_regions_all})
are rigorously consistent with the HST observations between January and May 2010.
In agreement with our result, \citet{kleyna-hainaut2013} find that the two arcs must be the products of sheet- or line-like ejection. We cannot exclude that these lines are specially enhanced regions on an ejecta cone, as suggested by \citet{kleyna-hainaut2013}. However, this perspective requires several very specific assumptions on the velocity-direction relation, and fails to explain the intersection of the two arcs seen in the early images. Therefore, while we cannot exclude an impact as the cause for the dust ejection from P/2010 A2, we believe that rotational break-up may be the simpler explanation.
We note that the existence of the eastern trail in October 2012 (cf. Fig.~\ref{fig:keck}) rules out the solution having $\theta_{A2}=74^\circ$ in \citet{kleyna-hainaut2013}, but not the one with $\theta_{A2}=0^\circ$.

\subsection{Rotational Breakup Interpretation}
\label{sec:interpretation}

Small asteroids are especially susceptible to the action of torques induced by radiation forces (``YORP'' effect).  The timescale to reach rotational instability under the action of YORP torques is of order 1 Myr for a 1 km radius body at 2 AU, and varies in proportion to the square of the radius.  The tiny nucleus of P/2010 A2 would have a YORP timescale of $\sim$5000 yr, considerably shorter than the collisional lifetime \citep{marzari-rossi2011}.  Such a short timescale makes YORP disruption a natural process to examine in the context of mass loss from P/2010 A2 and from sub-kilometer asteroids, generally.

Rotational instability under the action of YORP torques has been proposed as the cause of a high observed abundance of small binary asteroids \citep{walsh-richardson2008}.  The presumption is that the primary body has been fractured into a large number of mechanically independent sub-units (``blocks''), which can move in response to the changing spin of the body.  The details by which a spinning rubble pile loses mass are under discussion. In one model, increasing angular momentum can lead to an adjustment in the shape of the body, an elongation in the equatorial plane, and to the launch of material from the tips of the body, where gravitational acceleration is weakest.  Departing material moves slowly, and is subject to additional torques from the rotating, elongated primary that, under some circumstances, may lead to orbit circularization and the formation of a satellite \citep{walsh-richardson2008}.  In another model, increasing centripetal forces lead to a bifurcation of the primary and to the immediate formation of a binary \citep{jacobson-scheeres2011}.  

Whatever the details, we expect that the disruption of the primary would result in the launch of a considerable abundance of dust and debris particles, resulting in a mass-loss event perhaps not unlike that observed in P/2010 A2.  It is natural that material ejected in this way would be largely confined to a plane corresponding to the equatorial plane of the primary, but perhaps broadened by gravitational scattering impulses on the way out.  The duration of mass loss events caused by spin-up is unclear, but we expect repeated episodes of impulsive mass shedding separated by longer intervals during which the spin builds up to the critical value.

We outline in the following a qualitative and speculative model that might explain the existence of the arcs. In this model, the equatorial plane of the main nucleus is given by the plane of our best fits, and is perpendicular to the orbital plane. We assume that the nucleus rotates at a rate such that the rotational speed at the surface at the equator is higher than the escape speed. If the rotation speed at the surface is considerably higher than the escape speed, the material will leave on an almost straight trajectory, tangential to the surface. If the surface speed is only slightly higher than the escape speed, the trajectory will be bent around the nucleus under the action of gravity. An arc as observed in Fig.~\ref{fig:best_fit_empirical} can be formed by ejecting material at a range of speeds during a time interval short compared to the rotation period. 
Such a range of speeds could be due to either emission from different latitudes, or to speeds decreasing with time as a consequence of decelerating rotation.
The critical rotation period for a spherical body of density 3000~kg/m$^3$ is about 80 minutes, and longer for an elongated body. 

In the latitude model, the nucleus ejecting arc A would be rotating counterclockwise (seen from Earth), and material in region A3 would originate from close to the equator, while material in regions A1 and A2 would stem from latitudes that rotate just fast enough to overcome the nucleus gravity. The trajectories of the (slower) material in A1/A2 would be bent around the nucleus more strongly than in A3, thus forming the arc-shaped ejection pattern. The active latitude range would have to be rather narrow, because emission at high latitudes would give the material a significant component perpendicular to the equatorial plane, which we do not observe. 

In the decelerating model, all material could stem from the equatorial region. Material in region A3 would have been ejected first, while material in A1 and A2 would have been ejected just before the rotation period of the nucleus became sub-critical as a consequence of the dust ejection. Also in this case, the nucleus would be rotating counterclockwise, and arc ``A'' could have formed either through bending of the slow-particle trajectories, or due to the rotation itself, or a combination of both.

Arc ``B'' could be interpreted as being ejected in a similar manner from a second, fast-spinning source. The second source could have been a large fragment separated from the primary nucleus just before or just after the break-up that created arc A. 

\subsection{The Future}
P/2010 A2 was discovered long (9 or 10 months) after the disruption event, leading to the present difficulties in distinguishing between origin by impact and origin by rotational breakup.  The complexities of this study, and of that by \citet{kleyna-hainaut2013}, provide ample reason to hope that the next comparable asteroid disruption will be discovered much closer in time to its originating event.  Then, a simple observational test of formation scenarios will be possible.  Hypervelocity impacts invariably generate a fast component that will dissipate on timescales of weeks, as was observed in the impact-driven coma of large asteroid (596) Scheila \citep{bodewits-kelley2011, jewitt-weaver2011, ishiguro-hanayama2011a, ishiguro-hanayama2011b}. Rotational break-up of a fragmented body, on the other hand, cannot produce ejecta travelling much faster than the nucleus escape speed, even at the moment of origin.  Early-time observations will therefore provide a definitive discriminant between these models by showing the presence or absence of fast ejecta.  A separate consistency check based on nucleus rotation is possible, in principle, but difficult, in practice.  If rotational break-up is responsible then the central nucleus \textit{must} be in rapid (and probably excited) rotation while if impact is responsible the nucleus would not necessarily be a rapid rotator.  Determination of a slowly-spinning primary, therefore, would favor impact over rotational breakup.

\section{Summary}
\label{sec:summary}

We have re-examined a high resolution image sequence of P/2010 A2 obtained from the Hubble Space Telescope in 2010 and previously reported in \citet{jewitt-weaver2010}. Our method is to identify and follow the sky-plane motions of discrete structures in the dust tail of this object from 2010 January to May.  Using a dynamical model to account for the effects of Solar gravity and radiation pressure, we infer constraints on the ejection velocities of the discrete structures. 

\begin{itemize}

\item We find that the data are most simply described if the fragments were ejected at speeds comparable to the nucleus gravitational escape speed, in a common plane that intersects the nucleus.  

\item Planar emission is consistent with fragment ejection through rotational breakup of the parent nucleus but is less easily understood in the context of an impact origin.

\end{itemize}

\textbf{Acknowledgments}
Based on observations made with the NASA/ESA Hubble Space Telescope, obtained at the Space Telescope Science Institute, which is operated by the Association of Universities for Research in Astronomy, Inc., under NASA contract NAS 5-26555. These observations are associated with program number GO-12305.  DJ appreciates support from the NASA Planetary Astronomy program. We are grateful to Jan Kleyna for his comments to this manuscript.

{\it Facilities:} \facility{HST}

%\bibliographystyle{/home/agarwal/Latex/BibTex/astronat/apj/apj.bst}       % used style
%\bibliography{/home/agarwal/Latex/refs}  % reference entry

\clearpage

\begin{thebibliography}{18}
\expandafter\ifx\csname natexlab\endcsname\relax\def\natexlab#1{#1}\fi

\bibitem[{{Bodewits} {et~al.}(2011){Bodewits}, {Kelley}, {Li}, {Landsman},
  {Besse}, \& {A'Hearn}}]{bodewits-kelley2011}
{Bodewits}, D., {Kelley}, M.~S., {Li}, J.-Y., {et~al.} 2011, ApJL, 733, L3

\bibitem[{{Burns} {et~al.}(1979){Burns}, {Lamy}, \& {Soter}}]{burns-lamy1979}
{Burns}, J.~A., {Lamy}, P.~L., \& {Soter}, S. 1979, Icarus, 40, 1

\bibitem[{{Hainaut} {et~al.}(2012){Hainaut}, {Kleyna}, {Sarid}, {Hermalyn},
  {Zenn}, {Meech}, {Schultz}, {Hsieh}, {Trancho}, {Pittichov{\'a}}, \&
  {Yang}}]{hainaut-kleyna2012}
{Hainaut}, O.~R., {Kleyna}, J., {Sarid}, G., {et~al.} 2012, A\&A, 537, A69

\bibitem[{{Housen} \& {Holsapple}(2011)}]{housen-holsapple2011}
{Housen}, K.~R., \& {Holsapple}, K.~A. 2011, Icarus, 211, 856

\bibitem[{{Ishiguro} {et~al.}(2011{\natexlab{a}}){Ishiguro}, {Hanayama},
  {Hasegawa}, {Sarugaku}, {Watanabe}, {Fujiwara}, {Terada}, {Hsieh},
  {Vaubaillon}, {Kawai}, {Yanagisawa}, {Kuroda}, {Miyaji}, {Fukushima}, {Ohta},
  {Hamanowa}, {Kim}, {Pyo}, \& {Nakamura}}]{ishiguro-hanayama2011b}
{Ishiguro}, M., {Hanayama}, H., {Hasegawa}, S., {et~al.} 2011{\natexlab{a}},
  ApJL, 741, L24

\bibitem[{{Ishiguro} {et~al.}(2011{\natexlab{b}}){Ishiguro}, {Hanayama},
  {Hasegawa}, {Sarugaku}, {Watanabe}, {Fujiwara}, {Terada}, {Hsieh},
  {Vaubaillon}, {Kawai}, {Yanagisawa}, {Kuroda}, {Miyaji}, {Fukushima}, {Ohta},
  {Hamanowa}, {Kim}, {Pyo}, \& {Nakamura}}]{ishiguro-hanayama2011a}
---. 2011{\natexlab{b}}, ApJL, 740, L11

\bibitem[{{Jacobson} \& {Scheeres}(2011)}]{jacobson-scheeres2011}
{Jacobson}, S.~A., \& {Scheeres}, D.~J. 2011, \icarus, 214, 161

\bibitem[{{Jewitt} {et~al.}(2013){Jewitt}, {Ishiguro}, \&
  {Agarwal}}]{jewitt-ishiguro2013}
{Jewitt}, D., {Ishiguro}, M., \& {Agarwal}, J. 2013, ApJL, 764, L5

\bibitem[{{Jewitt} {et~al.}(2011{\natexlab{a}}){Jewitt}, {Stuart}, \&
  {Li}}]{jewitt-stuart2011}
{Jewitt}, D., {Stuart}, J.~S., \& {Li}, J. 2011{\natexlab{a}}, AJ, 142, 28

%\bibitem[{{Jewitt} {et~al.}(2010){Jewitt}, {Weaver}, {Agarwal}, {Mutchler}, \&
%  {Drahus}}]{jewitt-weaver2010}
%{Jewitt}, D., {Weaver}, H., {Agarwal}, J., {Mutchler}, M., \& {Drahus}, M.
%  2010, Nature, 467, 817

\bibitem[{Paper 1}()]{jewitt-weaver2010}
{Jewitt}, D., {Weaver}, H., {Agarwal}, J., {Mutchler}, M., \& {Drahus}, M.
  2010, Nature, 467, 817

\bibitem[{{Jewitt} {et~al.}(2011{\natexlab{b}}){Jewitt}, {Weaver}, {Mutchler},
  {Larson}, \& {Agarwal}}]{jewitt-weaver2011}
{Jewitt}, D., {Weaver}, H., {Mutchler}, M., {Larson}, S., \& {Agarwal}, J.
  2011{\natexlab{b}}, ApJL, 733, L4

\bibitem[{{Kadota} {et~al.}(2010){Kadota}, {Blythe}, {Spitz}, {Brungard},
  {Paige}, {Festler}, {Beshore}, {Ahern}, {Boattini}, {Gibbs}, {Grauer},
  {Hill}, {Kowalski}, {Larson}, {Ryan}, {Sato}, {Birtwhistle}, \&
  {Marsden}}]{kadota-blythe2010}
{Kadota}, K., {Blythe}, M., {Spitz}, G., {et~al.} 2010, Minor Planet Electronic
  Circulars, 32

\bibitem[{{Kim} {et~al.}(2012){Kim}, {Ishiguro}, {Hanayama}, {Hasegawa},
  {Usui}, {Yanagisawa}, {Sarugaku}, {Watanabe}, \&
  {Yoshida}}]{kim-ishiguro2012}
{Kim}, J., {Ishiguro}, M., {Hanayama}, H., {et~al.} 2012, ApJL, 746, L11

\bibitem[{{Kleyna} {et~al.}(2013){Kleyna}, {Hainaut}, \&
  {Meech}}]{kleyna-hainaut2013}
{Kleyna}, J., {Hainaut}, O.~R., \& {Meech}, K.~J. 2013, A\&A, 549, A13

\bibitem[{{M{\" u}ller} {et~al.}(2001){M{\" u}ller}, {Green}, \&
  {McBride}}]{mueller-green2001}
{M{\" u}ller}, M., {Green}, S.~F., \& {McBride}, N. 2001, in ESA SP-495:
  Meteoroids 2001 Conference, 47--54

\bibitem[{{Marzari} {et~al.}(2011){Marzari}, {Rossi}, \&
  {Scheeres}}]{marzari-rossi2011}
{Marzari}, F., {Rossi}, A., \& {Scheeres}, D.~J. 2011, \icarus, 214, 622

\bibitem[{{Snodgrass} {et~al.}(2010){Snodgrass}, {Tubiana}, {Vincent},
  {Sierks}, {Hviid}, {Moissl}, {Boehnhardt}, {Barbieri}, {Koschny}, {Lamy},
  {Rickman}, {Rodrigo}, {Carry}, {Lowry}, {Laird}, {Weissman}, {Fitzsimmons},
  {Marchi}, \& {OSIRIS Team}}]{snodgrass-tubiana2010}
{Snodgrass}, C., {Tubiana}, C., {Vincent}, J.-B., {et~al.} 2010, Nature, 467,
  814

\bibitem[{{Walsh} {et~al.}(2008){Walsh}, {Richardson}, \&
  {Michel}}]{walsh-richardson2008}
{Walsh}, K.~J., {Richardson}, D.~C., \& {Michel}, P. 2008, Nature, 454, 188

\end{thebibliography}

\clearpage

\begin{deluxetable}{lccccc}
\tabletypesize{\scriptsize}
\tablecaption{Dates and geometry of the Hubble Space Telescope observations (see also \citet{jewitt-weaver2010}). For our analysis we use all images but the last, due to low spatial resolution and SNR.}
\tablewidth{0pt}
\tablehead{
\colhead{UT Date} & \colhead{$R$ [AU]\tablenotemark{a}} & \colhead{$\Delta$ [AU]\tablenotemark{b}}   & \colhead{$\alpha$ [deg]\tablenotemark{c}} & \colhead{Scale [km]\tablenotemark{d}} & \colhead{Plane [deg]\tablenotemark{e}}  }
\startdata
2010-Jan-25 & 2.018 & 1.078 & 11.5 & 30.96 & -1.28\\
2010-Jan-29 & 2.019 & 1.099 & 13.5 & 31.56 & -0.94\\
2010-Feb-22 & 2.034 & 1.286 & 23.1 & 36.93 & 0.90\\
2010-Mar-12 & 2.047 & 1.473 & 27.0 & 42.30 & 1.82\\
2010-Apr-02 & 2.066 & 1.717 & 28.8 & 49.31 & 2.40\\
2010-Apr-19 & 2.083 & 1.922 & 28.7 & 55.20 & 2.55\\
2010-May-08 & 2.105 & 2.150 & 27.4 & 62.15 & 2.46\\
2010-May-29 & 2.130 & 2.393 & 25.0 & 69.18 & 2.13\\
\enddata

%% Text for table notes should follow after the \enddata but before
%% the \end{deluxetable}. Make sure there is at least one \tablenotemark
%% in the table for each \tablenotetext.
\tablenotetext{a}{Heliocentric distance in AU at the mid-time of the observatios.}
\tablenotetext{b}{Geocentric distance in AU at the mid-time of the observations.}
\tablenotetext{c}{Phase angle [degrees] at the mid-time of the observations.}
\tablenotetext{d}{Image scale, kilometers per 0.0396 arcsecond pixel.}
\tablenotetext{e}{Elevation of the Earth above the orbital plane of P/2010 A2.}
\label{tab:obs_geometry}
\end{deluxetable}

\clearpage

\begin{figure}
\epsscale{0.8}
\plotone{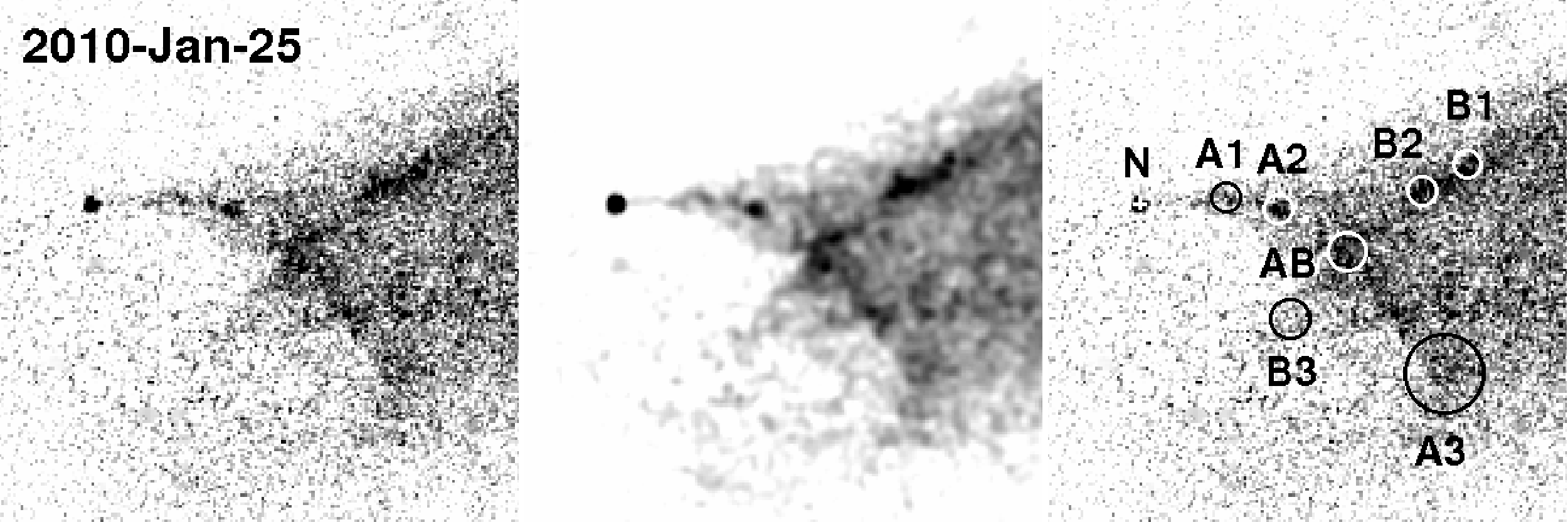}
\plotone{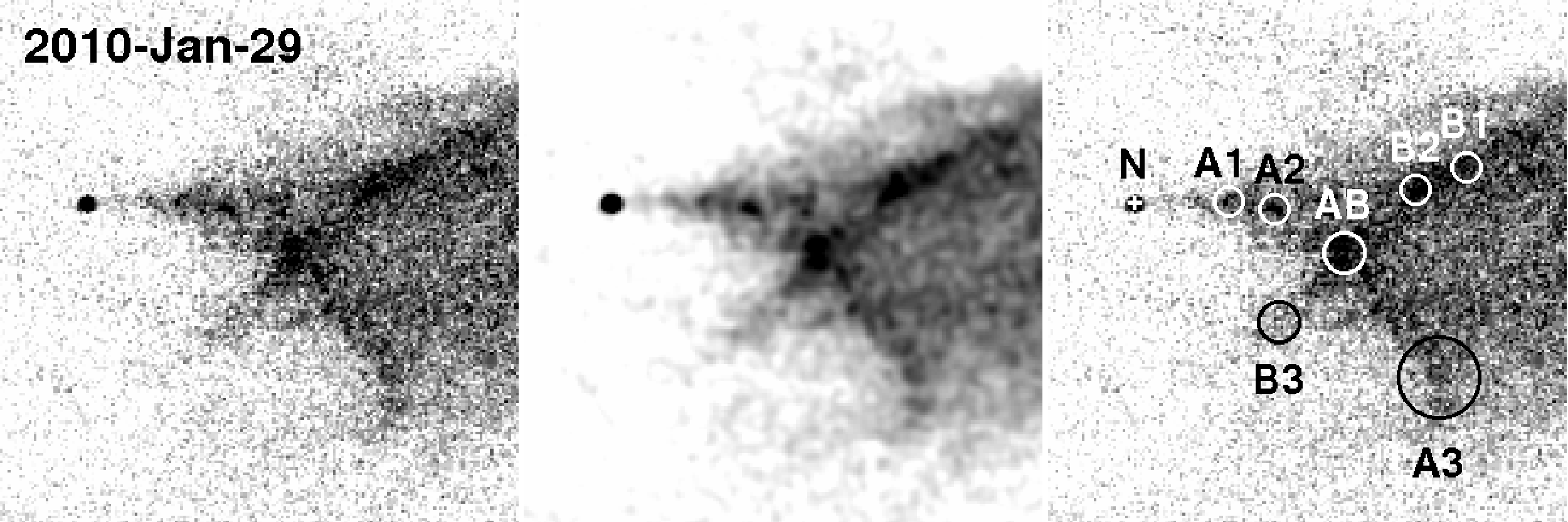}
\plotone{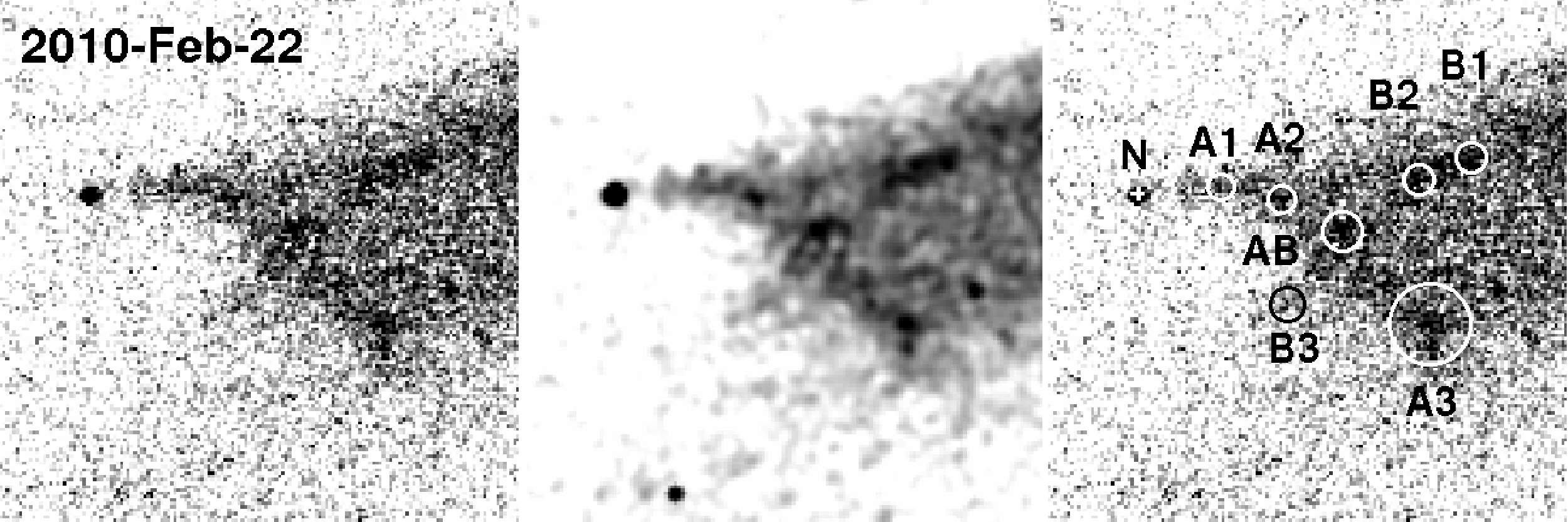}
\plotone{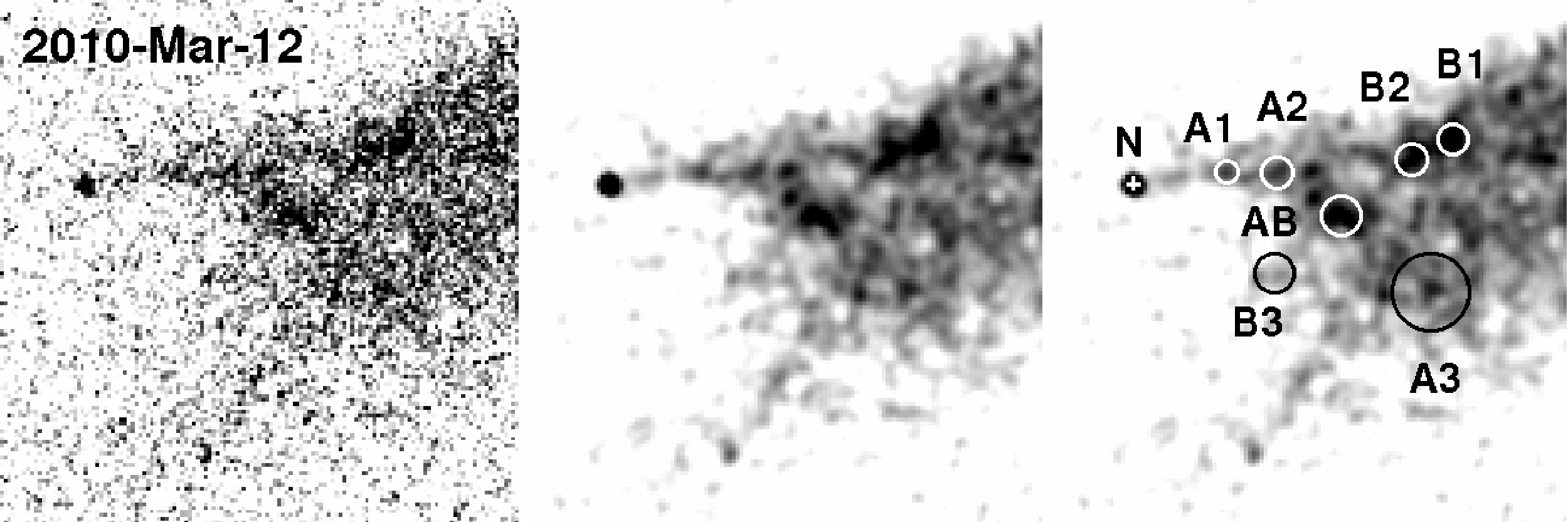}
\caption{See caption below second part.}
\end{figure}
\addtocounter{figure}{-1}

\begin{figure}
\epsscale{0.7}
\plotone{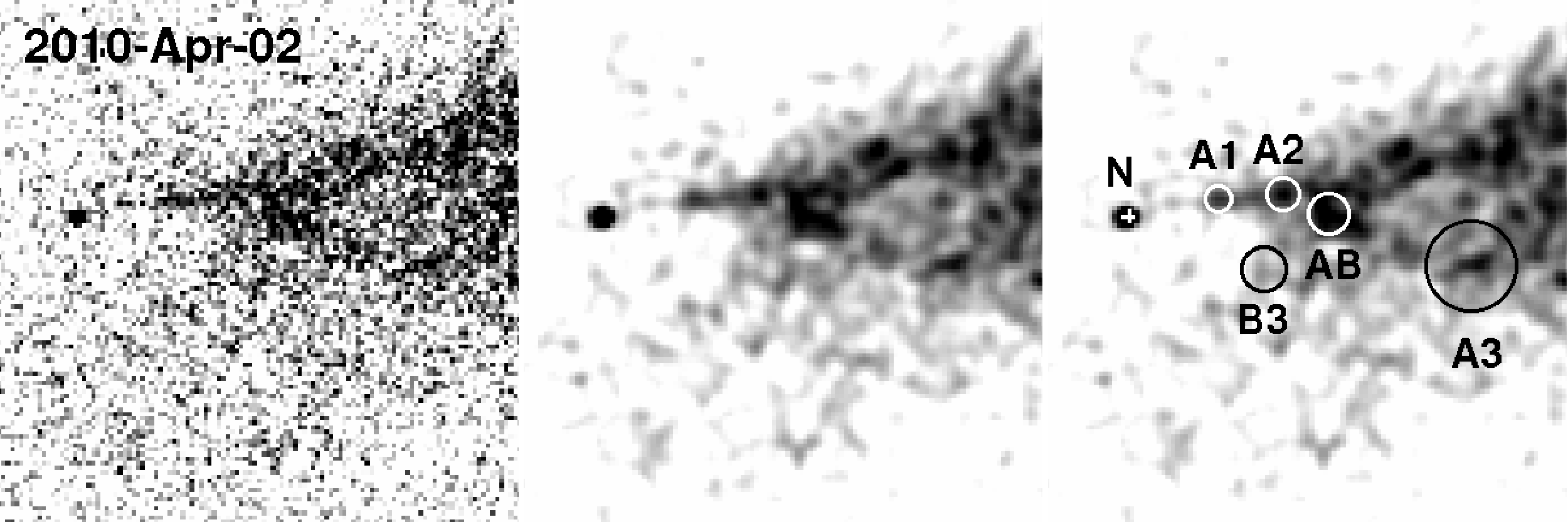}
\plotone{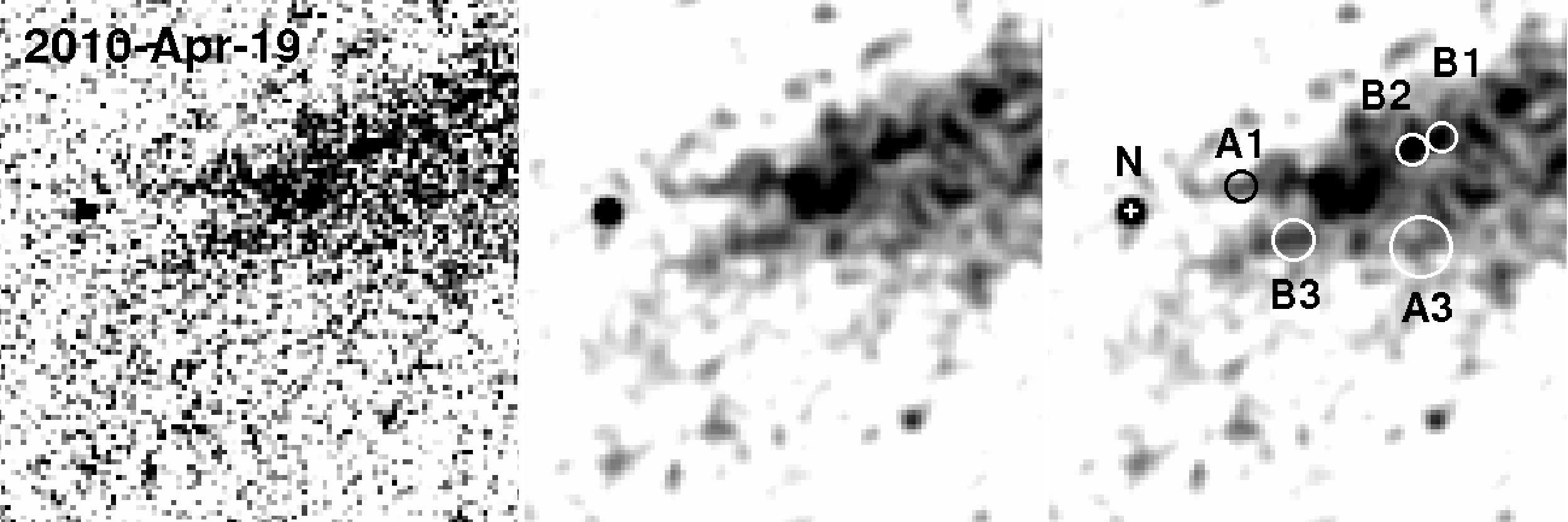}
\plotone{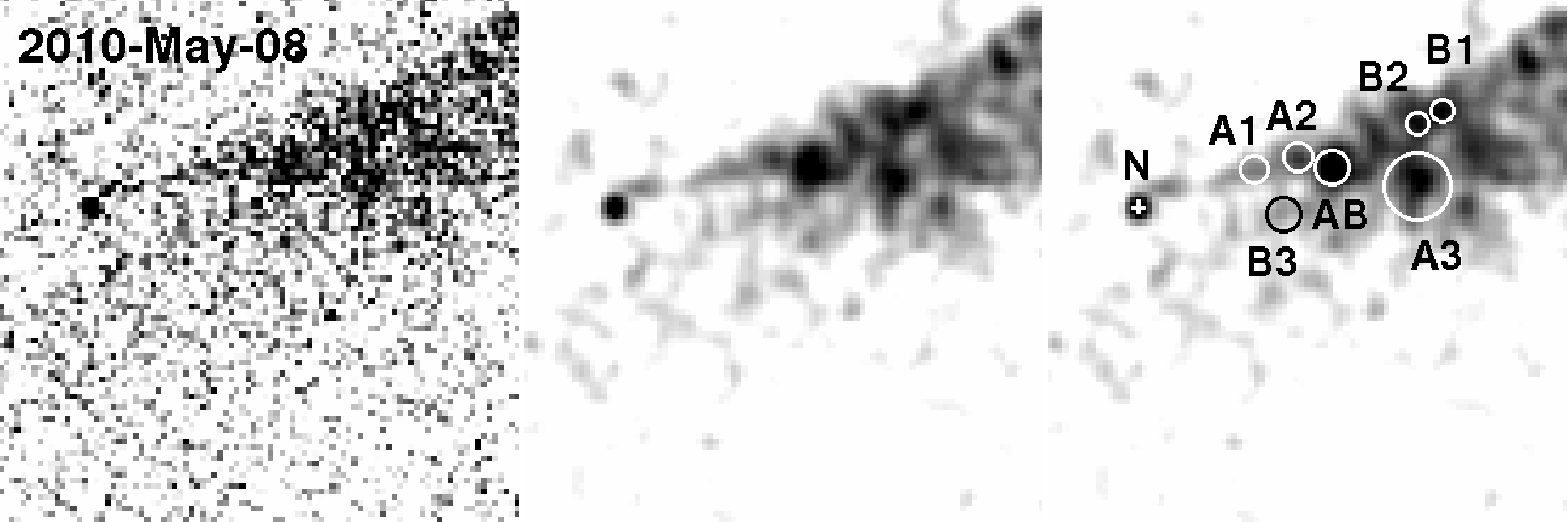}
\caption{Head of the tail of P/2010 A2 throughout the HST/WFC3 image series taken between 2010 January and May, showing the $X$-shaped, ribbon-like structure. 
Each image corresponds to about 7000\,km at the distance of the asteroid. Since the distance from Earth increased from 1.1\,AU in January to 2.1\,AU in May, the projected resolution deteriorates with time. The images have 0.04-arcsecond pixels and are combinations of exposures with total integration times of about 2,600s through the F606W filter. 
Left: original resolution, middle: smoothed with a Gaussian of 3 pixels radius, right: displaying the regions we study. 
We have selected regions at the ends of the cross and the intersection point of the two arms (circled and labelled). The northern ends of the arcs are marked by  elongated bright regions. We defined for each two separate circles marking the ends of these regions. For our analysis we assume that the material in each circular region remains the same throughout the image series. The circles are in black or white to maximise contrast.}
\label{fig:region_nomenclature}
\end{figure}

\begin{figure}
\epsscale{1.0}
\plotone{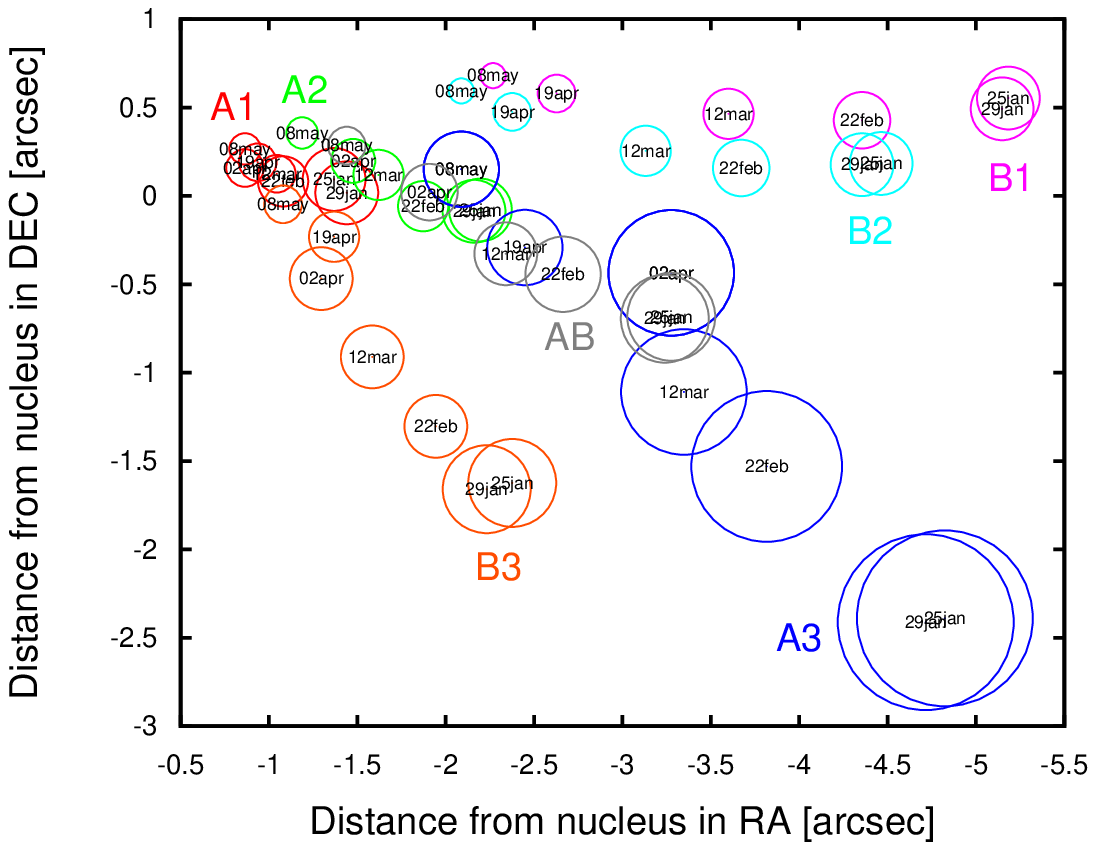}
\caption{Paths of the regions selected in Fig.~\ref{fig:region_nomenclature} through the HST image series in coordinates relative to the nucleus. The size of each circle corresponds to the size of the region we study, determined visually in Fig.~\ref{fig:region_nomenclature}.
We study a large number of test particles of different sizes and ejected to all directions from the nucleus. If a test particle falls into the circle of a specific region on all observation dates, we consider it a valid solution for that region. } 
\label{fig:region_paths}
\end{figure}

%\clearpage

\begin{figure}
\epsscale{0.7}
\plotone{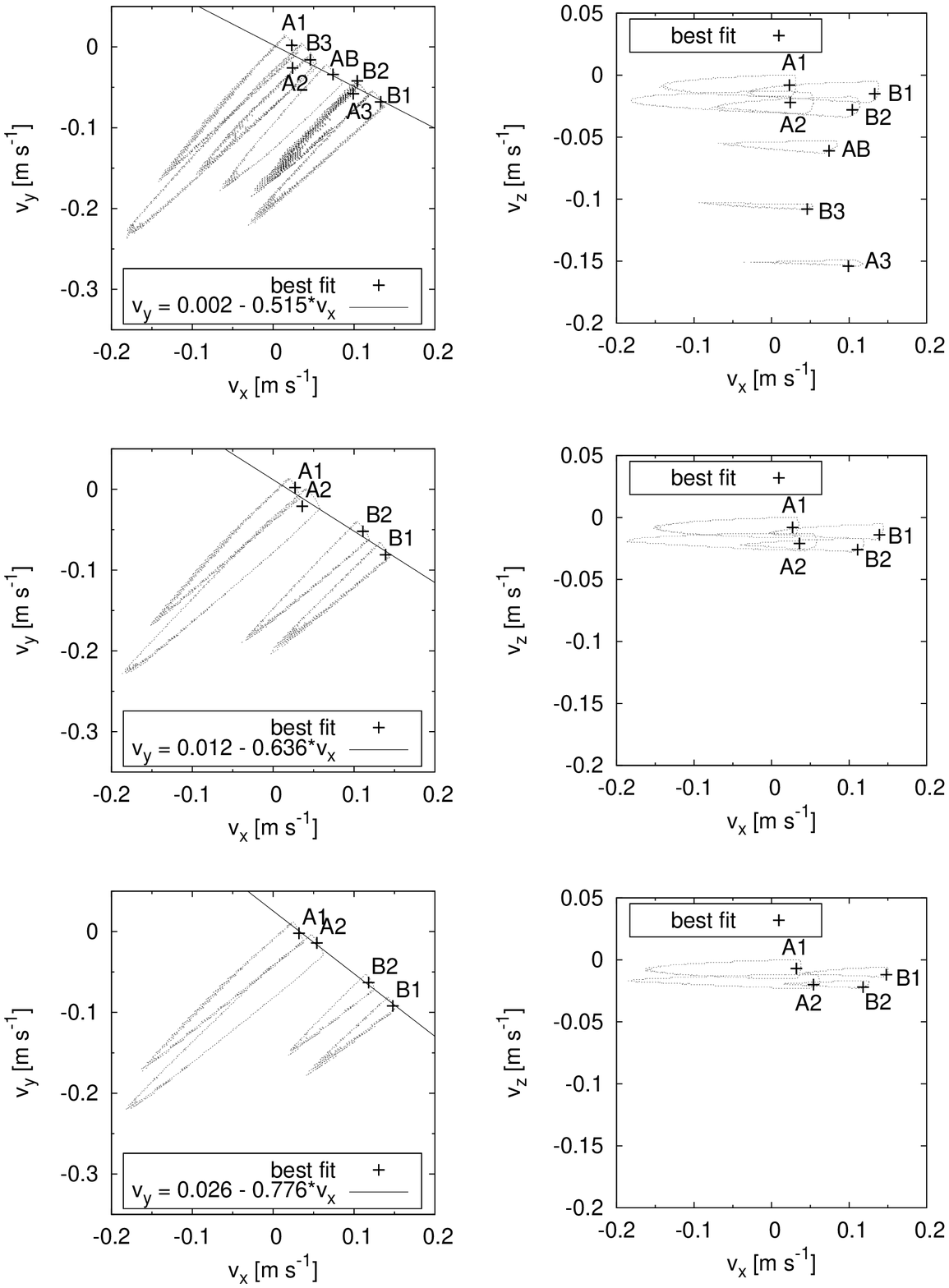}
\caption{Regions of possible solutions in velocity space for assumed emission dates on 2009 February 9 (top), March 2 (middle), and March 23 (bottom). Left: projection to the $v_x$-$v_y$-plane, right: projection to the $v_x$-$v_z$-plane. The contours outline the regions of allowable solutions (the test particle being inside the corresponding image region on all observation dates), while the best fitting solutions are marked by crosses. Solutions of similar fit quality lie on isolines having similar shapes as the contour shown. The straight line in the $v_x$-$v_y$-projection (left) is a linear fit to the positions of the best solutions.}
\label{fig:vspace_regions_all}
\end{figure}

\clearpage

\begin{figure}
\epsscale{1}
\plotone{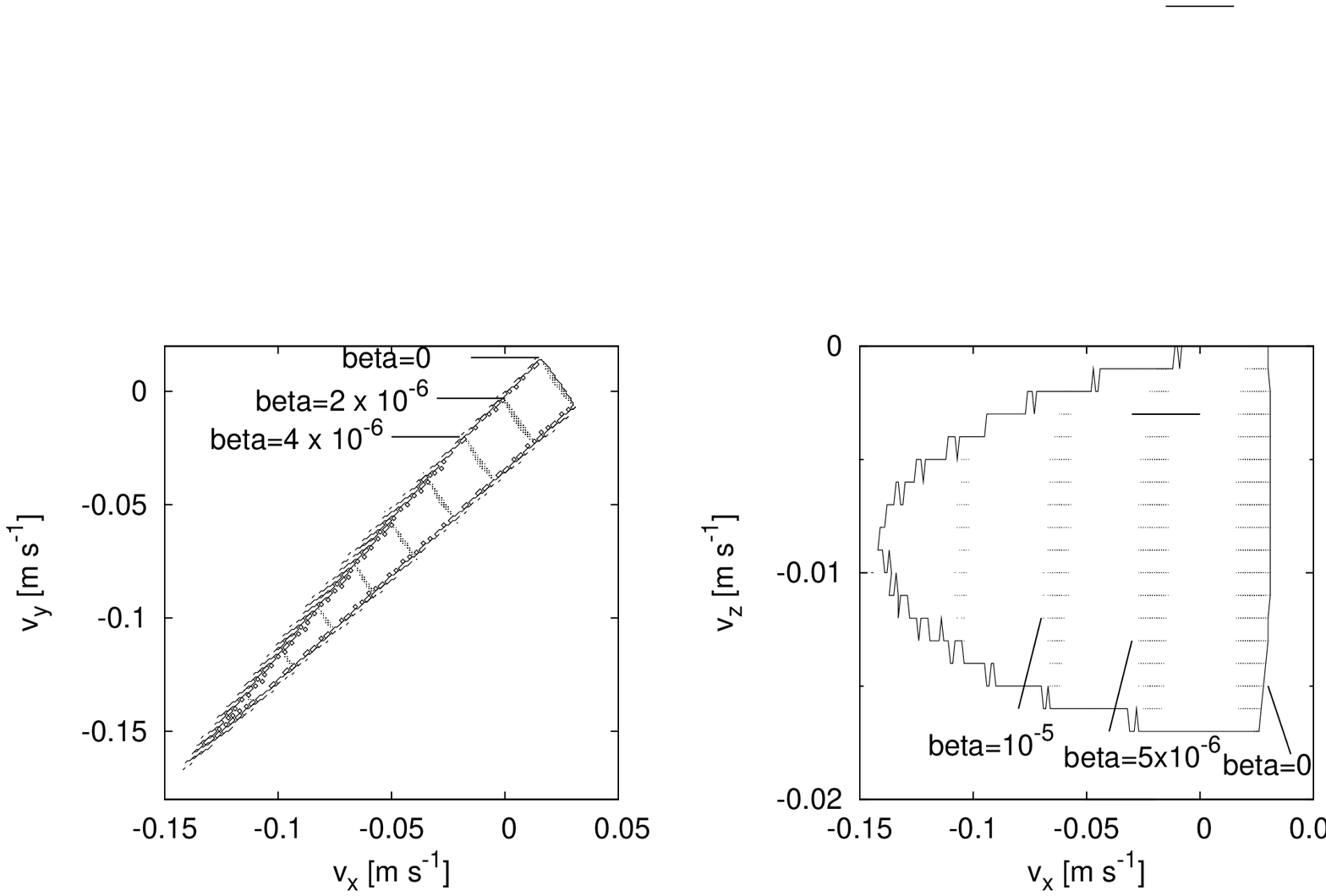}
\caption{Example for surfaces of constant radiation pressure parameter $\beta$ projected to the $v_x$-$v_y$-plane (left) and $v_x$-$v_z$-plane (right): for a given value of $\beta$, all allowable solutions for a specific region lie close to a plane that is perpendicular to the $v_x$-$v_y$-plane. Planes of constant $\beta$ are parallel to each other, but not parallel to the plane of best fits. The linear relation between ejection velocity and $\beta$ is described by Eq.~\ref{eq:relation_v_beta}.}
\label{fig:beta_sheets_example}
\end{figure}

\clearpage

\begin{figure}
\epsscale{0.7}
\plotone{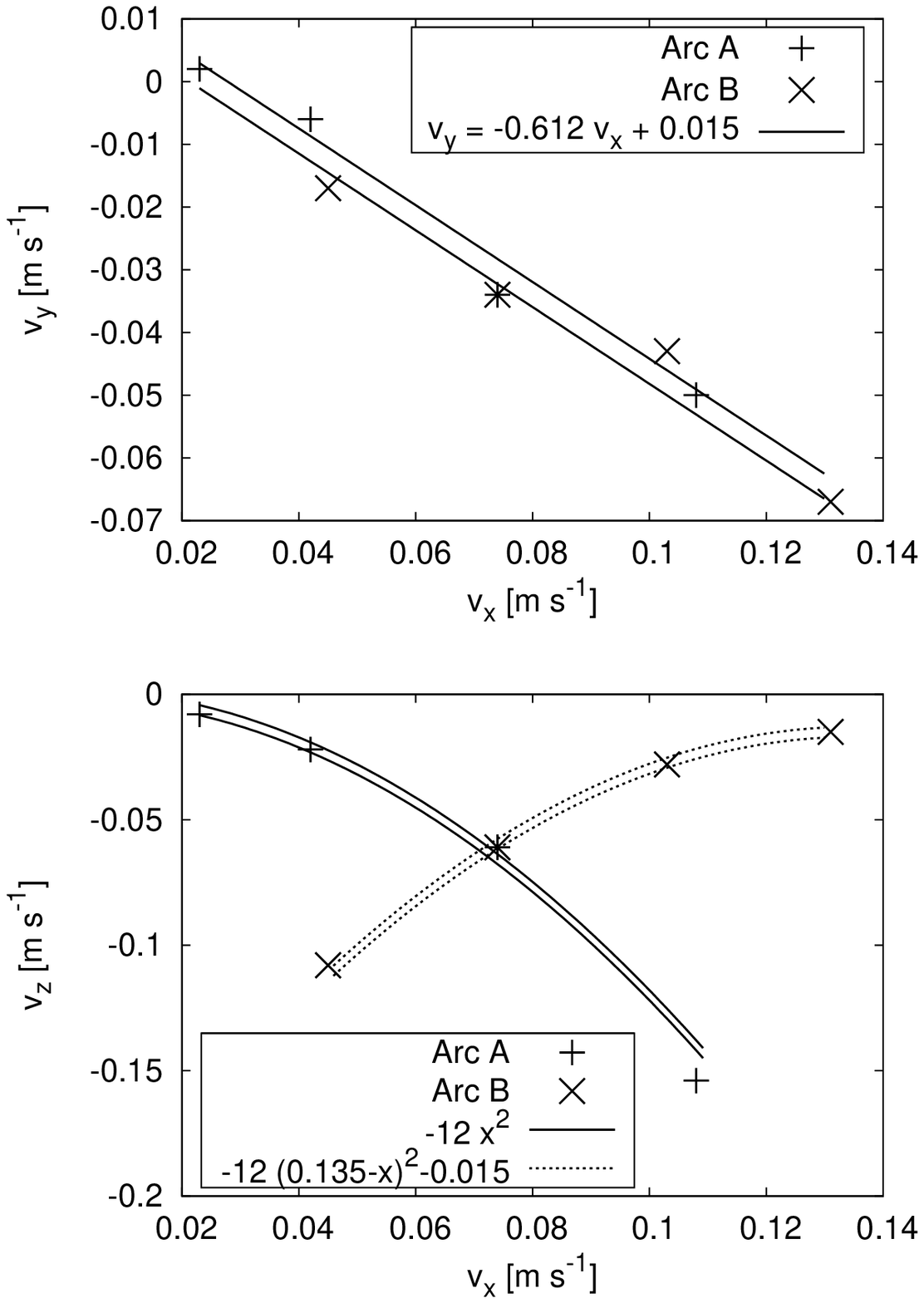}
\caption{Solutions with minimum $D_{j,k}$ for $\beta=10^{-7}$. ``+'' marks solutions for regions on arc A, ``x'' for arc B. Top: projection to the $v_x$-$v_y$-plane, bottom: projection to the $v_x$-$v_z$-plane. We fitted empirical relations to the coordinates of these solutions in order to produce a model image with a smooth distribution of initial velocities. The relation between $v_x$ and $v_y$ is similar for both arcs and can be described by a straight line (left). To describe $v_z (v_x)$ (right), we use quadratic functions, specific to each arc.}
\label{fig:best_fit_empirical}
\end{figure}

\clearpage

\begin{figure}
\epsscale{1.0}
\plotone{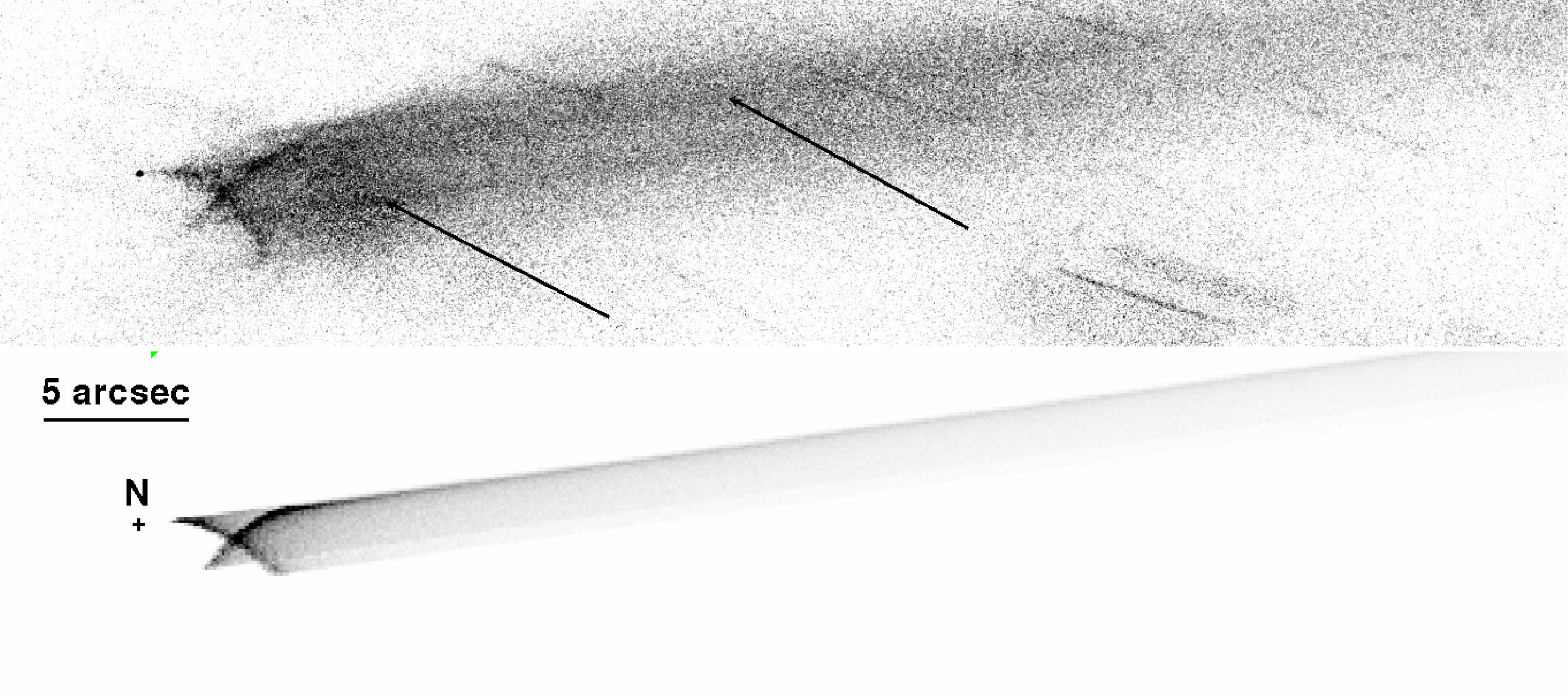}
\caption{Top: HST image of P2010/A2 on 2010 January 29. Bottom: Simulated image for the same date and with the same scale, based on the best-fitting ejection velocities derived in Sec.~\ref{sec:results}. This image was made by calculating the positions of particles ejected on 2009 February 9 with random velocities as described in Fig.~\ref{fig:best_fit_empirical}, and radiation pressure parameters $\beta$ corresponding to a power-law differential size distribution with an exponent of -3.3. Since the model image is based on our study of discrete regions in the $X$, it does not reproduce the diffuse parts surrounding it. Note how the filaments marked with arrows in the HST image are explained as the result of radiation pressure spreading of dust ejected towards the directions of region B3 and A1,A2,B1,B2 respectively.}
\label{fig:simimage}
\end{figure}

\clearpage
\begin{figure}
\epsscale{1}
\plotone{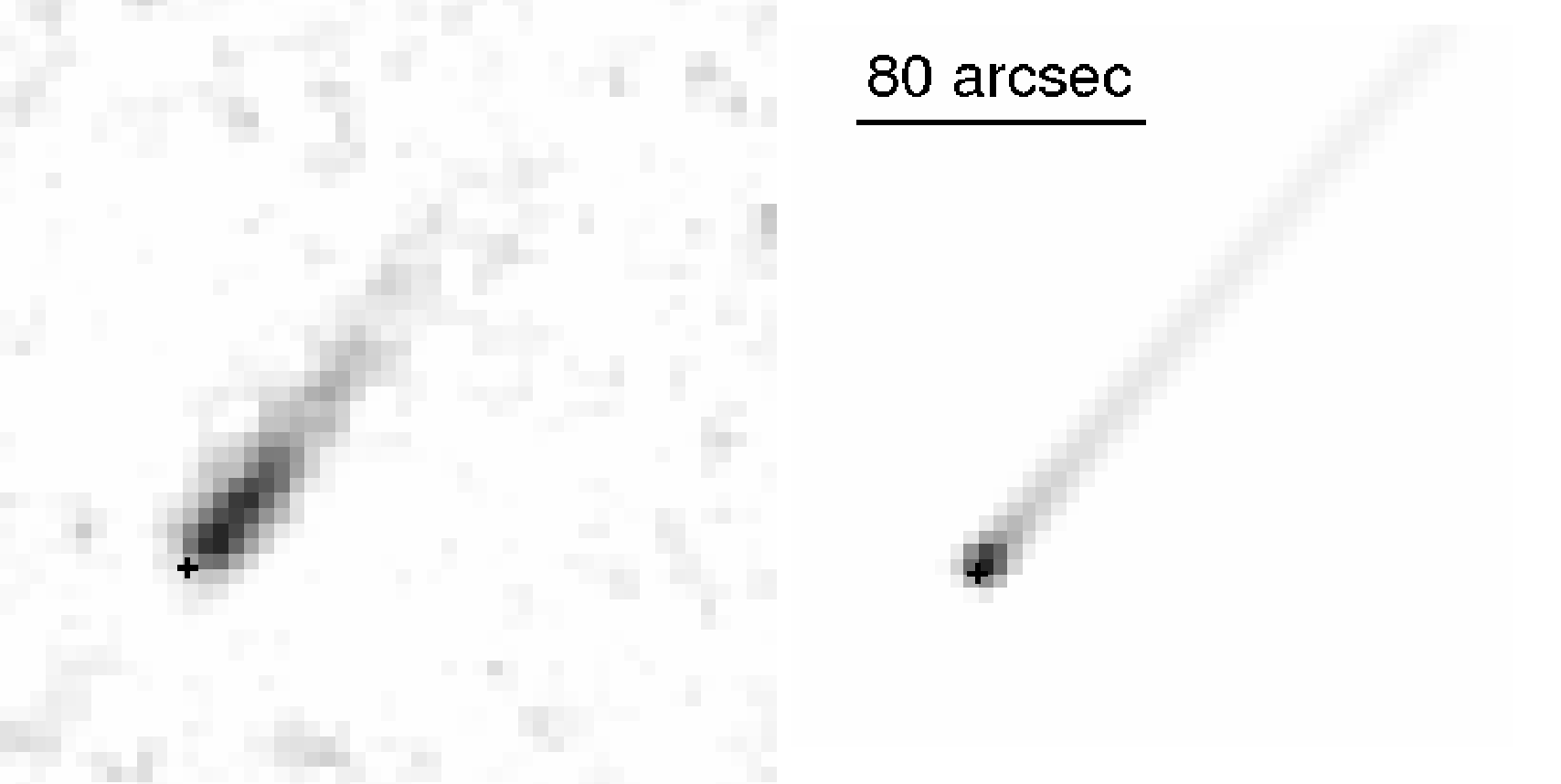}
\caption{Left: P/2010 A2 observed with the OSIRIS Narrow Angle Camera on board the Rosetta spacecraft on 16 March 2010 \citep{snodgrass-tubiana2010}. 
Right: Model image for the same date and observer position, generated with the parameters described in Fig.~\ref{fig:simimage} and smoothed with a Gaussian of radius 8 arcsec (2 pixels). The latter accounts for the combined effects of PSF and image stacking. In both panels, North is up and East is to the left. The crosses mark the position of the nucleus.}
\label{fig:ros}
\end{figure}

\clearpage

\begin{figure}
\epsscale{1}
\plotone{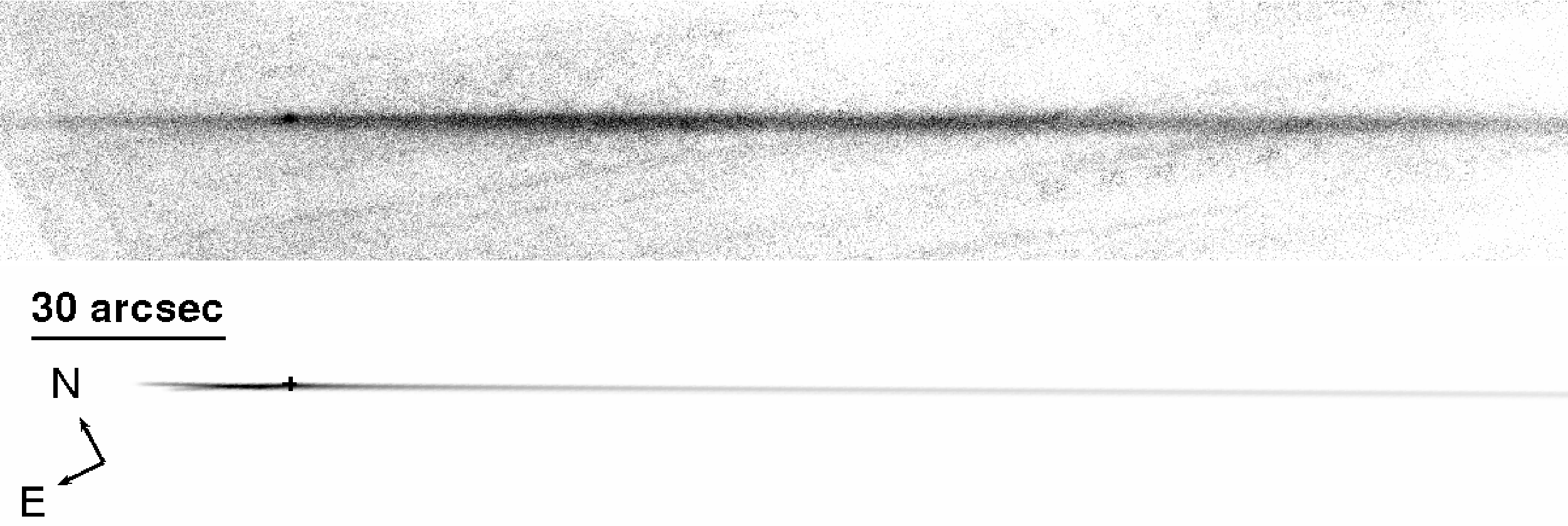}
\caption{Top: Keck observation of P/2010 A2 on 14 October 2012 \citep{jewitt-ishiguro2013}. Bottom: model image for the same date generated with the parameters used for Fig.~\ref{fig:simimage} and smoothed with a Gaussian of radius 0.5 arcsec to match the seeing at the time of the observation. In the model, the material forming the $X$ in 2010 is found to the east of the nucleus in 2012, consistent with the observation of the eastern trail in the top image. The $x$ structure cannot be discerned because of resolution and projection effects.}
\label{fig:keck}
\end{figure}

\end{document}